\documentclass[12pt]{article}

\usepackage{amssymb,amsmath,amsfonts,eurosym,geometry,ulem,graphicx,caption,color,setspace,sectsty,comment,footmisc,caption,pdflscape,subfigure,array,hyperref, placeins}
\usepackage[square,numbers]{natbib}
\usepackage{textcomp}

\usepackage{lineno}
\usepackage{pythonhighlight}
\usepackage{hyperref}
\usepackage{enumitem}
\usepackage{placeins}
\usepackage{amsmath}
\usepackage{booktabs} % For better horizontal lines
\usepackage{siunitx}  % For aligning decimal numbers

\title{Noisereduce: Domain General Noise Reduction for Time Series Signals}
\author{Tim Sainburg$^{1,2,3^\ast}$, Asaf Zorea$^{4,^\ast}$}

\begin{document}

\maketitle
\noindent
\scriptsize{*equal contribution}\\
\scriptsize{$^{1}$Department of Neurobiology, Harvard Medical School}\\
\scriptsize{$^{2}$Department of Molecular and Cellular Biology, Harvard University}\\
\scriptsize{$^{3}$Department of Organismic and Evolutionary Biology, Harvard University}\\
\scriptsize{$^{4}$Independent Researcher}\\

\begin{abstract}

Extracting signals from noisy backgrounds is a fundamental problem in signal processing across a variety of domains.
In this paper, we introduce Noisereduce, an algorithm for minimizing noise across a variety of domains, including speech, bioacoustics, neurophysiology, and seismology. 
Noisereduce uses spectral gating to estimate a frequency-domain mask that effectively separates signals from noise. It is fast, lightweight, requires no training data, and handles both stationary and non-stationary noise, making it both a versatile tool and a convenient baseline for comparison with domain-specific applications. We provide a detailed overview of Noisereduce and evaluate its performance on a variety of time-domain signals.
% we detail Noisereduce and evaluate its performance on a variety of time-domain signals.
\end{abstract}

% parallel computing

\section{Introduction}

Natural signals such as speech, electrophysiology, and bioacoustics are challenging to record in isolation. Sensors, both biological and artificial, tend to record these signals in the context of noisy environments. To record from a singing songbird in its natural environment, for example, a microphone will pick up not only the bird but the richness of its sensory environment—a babbling brook, chirping crickets, wind passing through leaves, and the croaks of a nearby frog. Such 'noise' can both provide important context for the signal of interest and important confounds. For example, a classifier trained to predict bird species from song recordings can be biased by environmental context; a babbling brook in the background might cause a Wood Thresh to be classified as a Robin. That error would in turn lead to downstream inaccuracies in estimating the migratory patterns of both birds. These same technical challenges arise in a variety of domains, from detecting action potentials to distinguishing seismic events from human activity.

Determining what constitutes noise versus signal is highly context-dependent. Consider two researchers: one focusing on the croaking of the American Bullfrog and the other analyzing the song of the Wood Thrush. They might approach the same audio recording yet define signal and noise in vastly different ways. Fortunately for these hypothetical researchers, the vocalizations of Bullfrogs and Wood Thrushes can be relatively easily distinguished from each other. Bullfrogs produce sounds in a frequency range of approximately 200-2000 Hz \cite{megela2008analyzing}, whereas Wood Thrushes vocalize in a higher spectrum, roughly 2000-9000 Hz \cite{injaian2021aircraft}. Therefore, by applying a simple low-pass or high-pass filter, each researcher can effectively isolate the vocalizations of their respective species with minimal effort.

Signal and noise events that overlap spectro-temporally pose a greater challenge but are not insurmountable. If the signal and noise retain identifiable structures, we can devise algorithms to exploit these structures and eliminate unwanted noise. For instance, the persistent 60-Hz hum from nearby electronics in a poorly grounded electrophysiology implant exhibits temporal structure. This constant hum can be identified and algorithmically removed from the signal. Noise reduction algorithms harness these structural differences to distinguish and separate noise from the signal.

Two broad classes of noise reduction algorithms exist. The first class is broadly described as conventional, while the other class of algorithms is machine-learning-based \cite{taha2018survey}. A crucial difference between conventional algorithms and more modern machine-learning-based algorithms is the characteristic reliance of machine learning on large, often labeled, datasets. Such datasets do not exist in all domains and require machine learning expertise to reimplement in new domains. These algorithms have been exhaustively reviewed for various signal domains in prior literature \cite{mehrish2023review, taha2018survey, xie2021bioacoustic}. Here, we survey the utility of our algorithm, Noisereduce, a fast, domain-general, spectral-subtraction-based algorithm available in Python. Noisereduce has already been available open-source for over five years and has found utility in a variety of different domains including bioacoustics \cite{sainburg2020finding, sainburg2021toward, mcewen2023automatic, mcginn2023feature, michaud2023unsupervised, fleishman2023ecological}, brain-machine interfacing \cite{chen2024neural, lee2023towards, lee2023speech}, livestock welfare monitoring \cite{jung2021deep, bhatt2022experimental}, human emotion analysis \cite{mazzocconi2023you, li2023global}, medical and clinical diagnostics \cite{liu2021machine, zhu2022automated, mandala2023enhanced, spiller2024enhancing, li2024multi}, seismic monitoring \cite{maher2024automated}, and many other domains. Until now, its performance has not been rigorously validated. Here, we address this by validating Noisereduce on several time-domain signals and comparing it to other conventional algorithms. Our findings show Noisereduce is fast and performs well, making it a strong candidate for domain-general applications where large datasets are unavailable and a solid baseline for comparing machine-learning-based algorithms.

\subsection{Noisereduce Algorithm}
Noisereduce is a form of spectral gating, or noise gating algorithm \cite{boll1979suppression}. 
Noise gates attenuate or suppress signals deemed noise, allowing the desired signal to pass through unaffected.
The spectral gate performs this gating in the spectro-temporal domain. Noisereduce accepts two inputs: (1) $X$, the time-domain recording to be denoised and (2, optionally) $X_{noise}$, a time-domain recording containing only noise, used to calculate noise statistics. Noisereduce operates through the following steps (Fig \ref{fig:algorithm}). 
\begin{enumerate}
  \item Estimate noise:
  \begin{enumerate}[label*=\arabic*.]
      \item Compute a Short-Time Fourier Transform (STFT; $S_n$) on each channel of the noise recording ($X_{noise}$).
      \item For each frequency channel, compute spectral statistics ($\mu_n$, $\sigma_n$) over the noise STFT ($S_n$).
      \item Compute a noise threshold based upon the statistics of the noise and the desired sensitivity. 
  \end{enumerate}
  \item Mask noise:
  \begin{enumerate}[label*=\arabic*.]
      \item Compute a STFT ($S_X$) over each channel of the recording ($X$).
      \item Compute a mask ($M$) over the signal STFT ($S_X$), based on the thresholds for each frequency channel. 
      \item (optional) smooth the mask ($M_{smooth}$) with a filter over frequency and time
      \item Apply the mask ($M_{smooth}$) to the STFT of the signal ($S_X$) to produce the masked STFT ($S_m$).
      \item Invert the masked STFT ($S_m$) back into the time-domain ($X_{denoised}$).
  \end{enumerate}
\end{enumerate}
If the noise recording is not provided to the algorithm, the noise statistics are computed on directly on the recording ($X_s$). A more detailed description of the algorithm and its parameters are given in  \ref{sec:implementation}.

\setcounter{figure}{0}  
\begin{figure}[!htbp]
\centering
\includegraphics[width=0.86\textwidth]{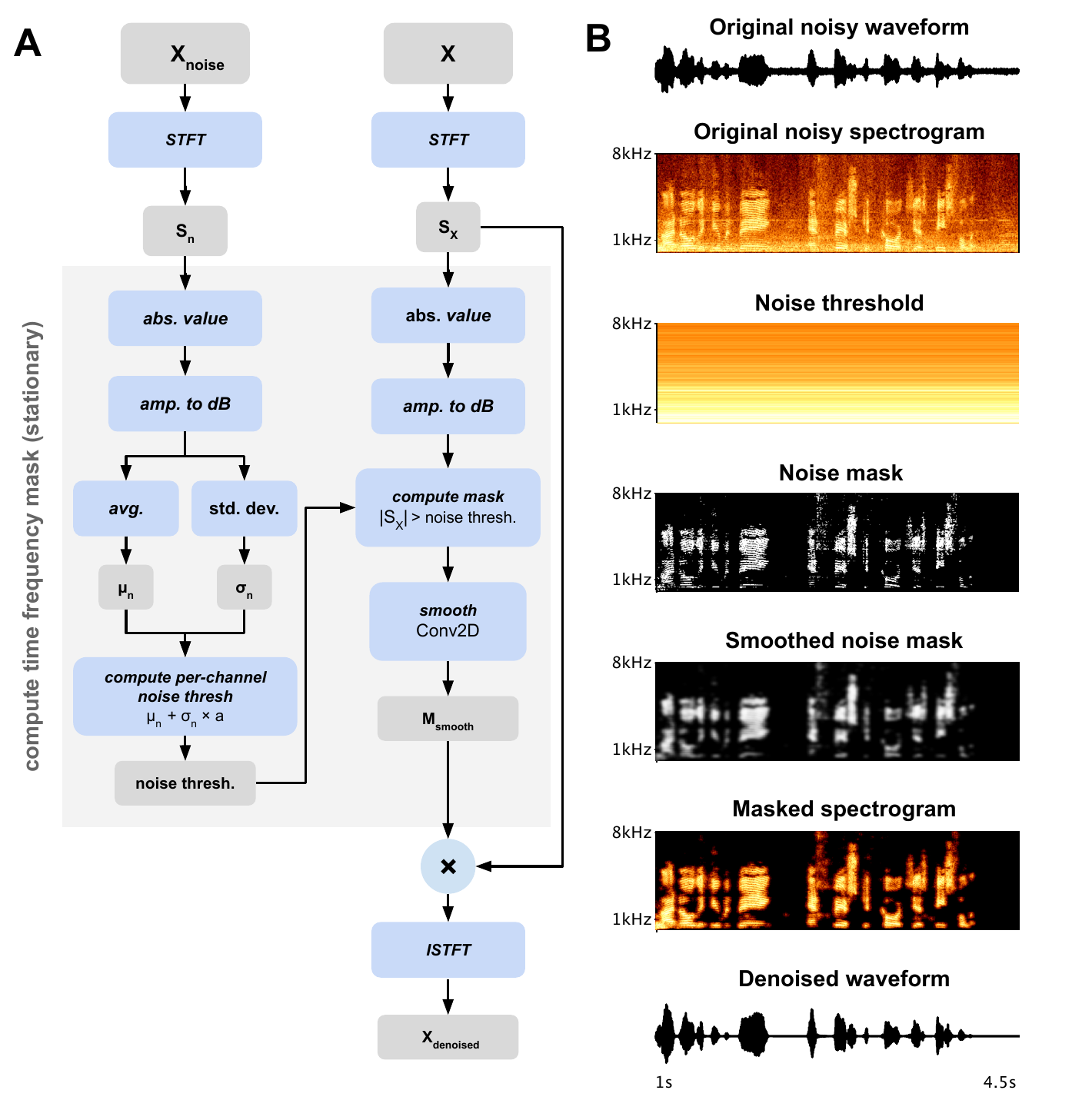}
\caption{
Basic outline of Noisereduce algorithm. (A) A block diagram of the steps of Noisereduce. The stationary version of the time-frequency mask is depicted. (B) An example waveform (U.S. President George W Bush stating "I know that human beings and fish can coexist peacefully") passing through the Noisereduce pipeline. The non-stationary algorithm is not shown here. 
}
\label{fig:algorithm}
\end{figure}
\FloatBarrier

\subsubsection{Nonstationary Noise Reduction}

In natural settings, background noise often varies over extended periods. For example, in bioacoustics, weather can shift within minutes, while in electrophysiology, the activity rates of nearby neurons may increase as animals transition between states, such as sleeping and waking. Consequently, it is advantageous to enable Noisereduce to adapt its noise definition over time \cite{sainburg2021toward}. To address this, we introduced a non-stationary variant of Noisereduce, where mask statistics are calculated using a sliding window across the signal rather than relying solely on an isolated noise clip. This non-stationary approach is particularly beneficial for signals such as those from hydrophones in underwater bioacoustics, where the engine hum of a boat can fluctuate as the hydrophone drifts toward and away from the boat towing it. Normalizing audio signals to channel-specific fluctuations in amplitude has proven useful for tasks like bioacoustic species identification \cite{lostanlen2018per}.

The non-stationary algorithm omits the need for a noise recording ($X_n$) since noise statistics are directly derived from the signal recording ($X_s$). In this revised approach, statistics for the noise threshold are computed over a sliding window for each frequency channel. This approach dynamically sets noise gate thresholds for each frequency channel, as opposed to static settings across the entire recording. For additional details, see Section \ref{sec:implementation}.

\begin{figure}[!htbp]
\centering
\includegraphics[width=1.0 \textwidth]{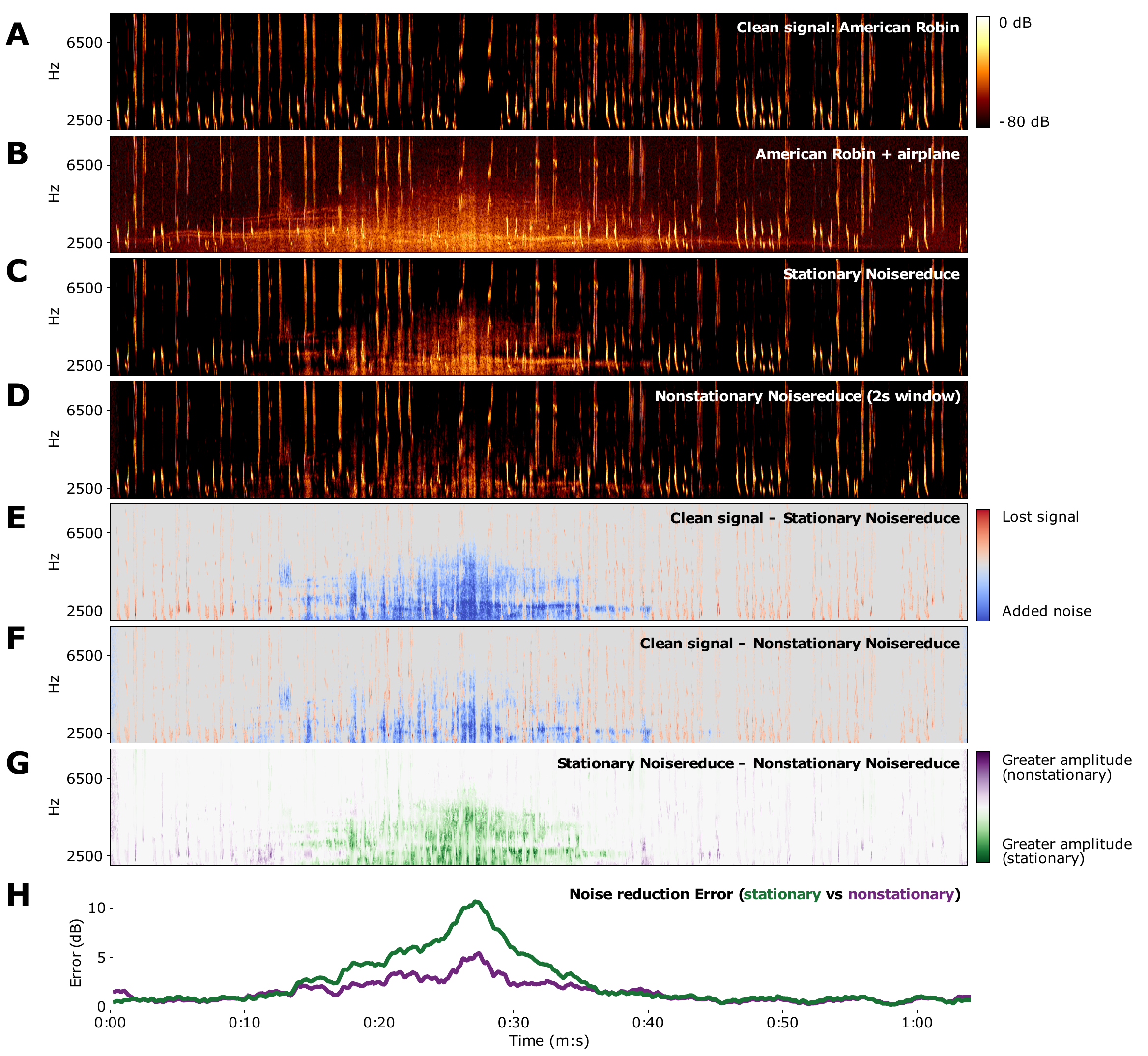}
\caption{
Comparison of stationary and non-stationary noise reduction. 
(A) Spectrogram of clean recording of an American Robin (Macaulay Library 321642131).
(B) Airplane noise imposed over Robin Recording.
(C-D) Denoising of (B) with (C) stationary noisereduce and (D)  nonstationary noisereduce (window size of 2 seconds)
(E-F) Magnitude error in stationary noisereduce vs ground truth for (E) stationary noisereduce and (F) nonstationary noisereduce. 
(G) Magnitude difference between stationary and nonstationary noisereduce. 
(H) Error (in dB) from ground truth for stationary (green) and nonstationary (purple) noisereduce. 
}
\label{fig:nonstationary}
\end{figure}

Figure \ref{fig:nonstationary} illustrates the non-stationary algorithm's utility. We took a one-minute recording of an American Robin (Macaulay Library 321642131; Fig \ref{fig:nonstationary}A) and added the non-stationary noise of an airplane passing overhead (Fig \ref{fig:nonstationary}B). We then applied both stationary and nonstationary Noisereduce (Fig \ref{fig:nonstationary}C-D). During the highest amplitude period of airplane noise, the stationary algorithm leaves additional noise artifacts in the recording, unlike the non-stationary version (Fig \ref{fig:nonstationary}E-G, Blue/Green, 20-30 seconds). Conversely, more of the signal is lost in sections with lower noise amplitude (e.g. 40-60 seconds, red/purple). We quantified this as the absolute error in dB, relative to the noise-free recording (Fig \ref{fig:nonstationary}H) exemplifting that the non-stationary algorithm performs consistently better with non-stationary noise in this case.

\FloatBarrier
\section{Experiments}
To evaluate the performance of Noisereduce, we tested it on a set of benchmark datasets across four domains: speech, bioacoustics, electrophysiology, and seismology (see \ref{sec:datasets}). We compared its results against several noise reduction algorithms (see \ref{sec:algorithms}). The evaluation metrics used in the comparison are detailed in \ref{sec:metrics}.

% To validate Noisereduce we compared it on a set of benchmark datasets across four domains: speech, bioacoustics, electrophysiology, and seismology (see \ref{sec:datasets}), against a series of other conventional, domain-general, noise reduction algorithms (see \ref{sec:algorithms}). The metrics used for the evaluation are summarized in \ref{sec:metrics}.

\subsection{Speech}
Speech is the best-established domain for enhancement and noise reduction \cite{loizou2013speech}. 
Many speech noise reduction applications are well-suited to machine learning methods, especially deep neural networks like convolutional neural networks (CNNs) \cite{8466892, macartney2018improvedspeechenhancementwaveunet}, long short-term memory networks (LSTMs) \cite{defossez2020realtimespeechenhancement, Hao_2021}, and Generative Adversarial Network (GANs) \cite{pascual2017seganspeechenhancementgenerative, Hao2019}, which outperform any conventional algorithm.
We therefore submit that Noisereduce in this domain for two purposes. First, as a candidate "conventional algorithm" baseline. Second, Noisereduce may remain useful for speech applications where machine-learning based approaches might not be well suited, such as out-of-domain speech signals, very lightweight applications where computational costs of machine-learning based approaches are too cumbersome, or in creating new datasets with varying manipulations on noise levels. 

\begin{figure}[!htbp]
\centering
\includegraphics[width=0.95 \textwidth]{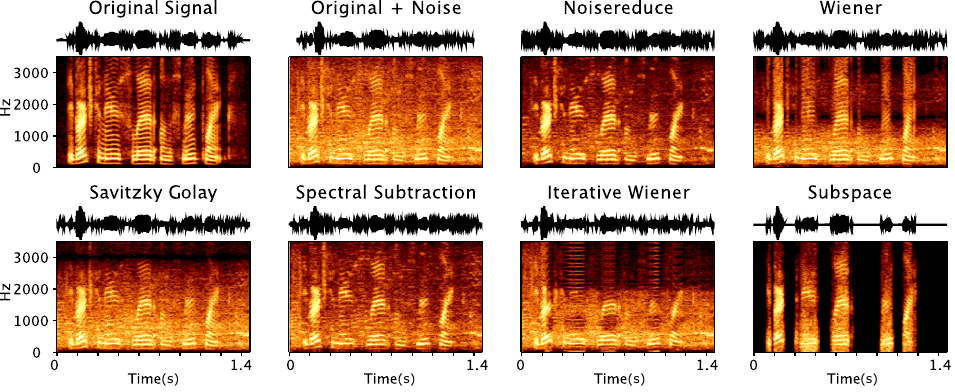}
\caption{
Noise reduction samples from different algorithms applied to the 'sp04' sample from the NOIZEUS dataset (SNR: 10 dB, exhibition noise).
}
\label{fig:speech}
\end{figure}

We evaluated Noisereduce against other noise reduction conventional algorithms on the NOIZEUS dataset  \cite{noauthor_noizeus:_nodate, yi_hu_subjective_2006, hu_evaluation_2008} across various SNR levels (0, 5, 10, and 15 dB). Examples of speech spectrograms obtained with Noisereduce, Wiener \cite{1163086}, Iterative Wiener \cite{1163086}, Subspace \cite{397090, 5743782}, Spectral Subtraction \cite{boll1979} and Savitzky-Golay \cite{savitzky1964} appear in Fig \ref{fig:speech}. Particularly, Noisereduce preserves the speech signal without distortions, unlike other algorithms that add artifacts, particularly under low SNR.
The performance metrics used were Short-Time Objective Intelligibility (STOI) \cite{taal_short-time_2010} and Perceptual Evaluation of Speech Quality (PESQ) \cite{941023}, which assess speech intelligibility and quality, respectively. STOI and PESQ results are in Tables \ref{tab:stoi} and \ref{tab:pesq}, showing that Noisereduce outperforms other conventional algorithms at all tested SNR levels.
%For instance, at 0 dB SNR, Noisereduce achieved a STOI score of $0.683 \pm 0.004$, while the next best result, the Baseline (unprocessed noisy signal), scored $0.671 \pm 0.004$ ($p < 0.05$). Similar  improvements were observed at higher SNRs levels ($p < 0.002$). When compared with other noise reduction algorithms, Noisereduce also demonstrated statistically significant improvements in STOI scores across all SNRs ($p < 0.001$).
%At 0 dB SNR, Noisereduce achieved a PESQ score of $1.559 \pm 0.008$, significantly higher than the Baseline's $1.421 \pm 0.009$ ($p < 0.001$). Similar improvements were observed at higher SNRs ($p < 0.001$).
%Comparisons with other noise reduction algorithms also showed that Noisereduce significantly outperformed them in PESQ scores across all SNR levels ($p < 0.001$).

% \FloatBarrier

\sisetup{
    detect-all,
    separate-uncertainty = true,
    table-format=1.3
}
\begin{table}[ht]
    \small
    \centering
    \begin{tabular}{
        l
        S[table-format=1.3(3)]
        S[table-format=1.3(3)]
        S[table-format=1.3(3)]
        S[table-format=1.3(3)]
    }
    \toprule
    {Algorithm} & {SNR 0} & {SNR 5} & {SNR 10} & {SNR 15} \\
    \midrule
    Baseline & 0.671 \pm 0.004 & 0.783 \pm 0.003 & 0.878 \pm 0.003 & 0.937 \pm 0.002 \\
    Iterative Wiener & 0.509 \pm 0.004 & 0.594 \pm 0.004 & 0.664 \pm 0.005 & 0.704 \pm 0.005 \\
    NoiseReduce (ours) & \textbf{0.683 $\pm$ 0.004} & \textbf{0.799 $\pm$ 0.003} & \textbf{0.893 $\pm$ 0.002} & \textbf{0.946 $\pm$ 0.002} \\
    Savitzky-Golay & 0.668 \pm 0.004 & 0.779 \pm 0.003 & 0.875 \pm 0.003 & 0.934 \pm 0.002 \\
    Spectral Subtraction & 0.417 \pm 0.003 & 0.451 \pm 0.002 & 0.479 \pm 0.002 & 0.493 \pm 0.002 \\
    Subspace & 0.608 \pm 0.004 & 0.682 \pm 0.003 & 0.712 \pm 0.004 & 0.724 \pm 0.004 \\
    Wiener & 0.668 \pm 0.004 & 0.766 \pm 0.004 & 0.840 \pm 0.003 & 0.879 \pm 0.002 \\
    \bottomrule
    \end{tabular}
    \caption{STOI performance metric on NOIZEUS dataset (mean $\pm$ SEM) for different algorithms across various SNR levels.}
    \label{tab:stoi}
\end{table}

\begin{table}[ht]
    \small
    \centering
    \begin{tabular}{
        l 
        S[table-format=1.3(3)]
        S[table-format=1.3(3)]
        S[table-format=1.3(3)]
        S[table-format=1.3(3)]
    }
    \toprule
    {Algorithm} & {SNR 0} & {SNR 5} & {SNR 10} & {SNR 15} \\
    \midrule
    Baseline & 1.421 \pm 0.009 & 1.600 \pm 0.010 & 1.878 \pm 0.011 & 2.238 \pm 0.013 \\
    Iterative Wiener & 1.374 \pm 0.010 & 1.516 \pm 0.011 & 1.687 \pm 0.013 & 1.874 \pm 0.017 \\
    Noisereduce (ours) & \textbf{1.559 $\pm$ 0.008} & \textbf{1.854 $\pm$ 0.009} & \textbf{2.286 $\pm$ 0.011} & \textbf{2.778 $\pm$ 0.012} \\
    Savitzky-Golay & 1.475 \pm 0.010 & 1.672 \pm 0.011 & 1.973 \pm 0.012 & 2.353 \pm 0.014 \\
    Spectral Subtraction & 1.493 \pm 0.010 & 1.733 \pm 0.009 & 2.064 \pm 0.011 & 2.449 \pm 0.012 \\
    Subspace & 1.415 \pm 0.009 & 1.407 \pm 0.008 & 1.380 \pm 0.006 & 1.379 \pm 0.007 \\
    Wiener & 1.458 \pm 0.009 & 1.634 \pm 0.009 & 1.858 \pm 0.011 & 2.095 \pm 0.012 \\
    \bottomrule
    \end{tabular}
    \caption{PESQ performance metric on NOIZEUS dataset (mean $\pm$ SEM) for different algorithms across various SNR levels.}
    \label{tab:pesq}
\end{table}

\FloatBarrier

To further evaluate its performance, we compared Noisereduce against a state-of-the-art deep learning-based model, Denoiser \cite{defossez2020realtimespeechenhancement}. While Denoiser had higher STOI and PESQ scores (Tables \ref{tab:denoiser_stoi} and \ref{tab:denoiser_pesq}), Noisereduce achieved competitive results with substantially lower computational overhead. Specifically, Denoiser requires over 33 million trainable parameters, whereas Noisereduce uses efficient signal processing techniques that require minimal computational resources and provide faster runtime (see \ref{sec:runtime}).

\begin{table}[ht]
    \small
    \centering
    \begin{tabular}{
        l 
        S[table-format=1.3(3)]
        S[table-format=1.3(3)]
        S[table-format=1.3(3)]
        S[table-format=1.3(3)]
    }
    \toprule
    {Algorithm} & {SNR 0} & {SNR 5} & {SNR 10} & {SNR 15} \\
    \midrule
    Noisereduce (ours) & 0.683 \pm 0.004 & 0.799 \pm 0.003 & 0.893 \pm 0.002 & 0.946 \pm 0.002 \\
    Denoiser & \textbf{0.796 $\pm$ 0.005} & \textbf{0.88 $\pm$ 0.003} & \textbf{0.927 $\pm$ 0.002} & \textbf{0.951 $\pm$ 0.002} \\
    \bottomrule
    \end{tabular}

    \caption{Comparison of STOI performance metric for Noisereduce and Denoiser across various SNR levels (mean $\pm$ SEM).}
    \label{tab:denoiser_stoi}
\end{table}

\begin{table}[ht]
    \small
    \centering
    \begin{tabular}{
        l 
        S[table-format=1.3(3)]
        S[table-format=1.3(3)]
        S[table-format=1.3(3)]
        S[table-format=1.3(3)]
    }
    \toprule
    {Algorithm} & {SNR 0} & {SNR 5} & {SNR 10} & {SNR 15} \\
    \midrule
    Noisereduce (ours) & 1.559 \pm 0.008 & 1.854 \pm 0.009 & 2.286 \pm 0.011 & \textbf{2.778 $\pm$ 0.012} \\
    Denoiser & \textbf{1.671 $\pm$ 0.015} & \textbf{2.04 $\pm$ 0.017} & \textbf{2.39 $\pm$ 0.018} & 2.703 \pm 0.023 \\
    \bottomrule
    \end{tabular}

    \caption{Comparison of PESQ performance metric for Noisereduce and Denoiser across various SNR levels (mean $\pm$ SEM).}
    \label{tab:denoiser_pesq}
\end{table}

\FloatBarrier

\subsection{Bioacoustics}
Bioacoustic signals are recorded across Earth's diverse bioregions, with conditions often unique to each dataset. Consequently, state-of-the-art machine-learning methods are rarely available \cite{xie2021bioacoustic}, making bioacoustics is an ideal domain for applying Noisereduce. 
To our knowledge, no benchmark dataset exists for bioacoustic noise reduction, unlike NOIZEUS \cite{noauthor_noizeus:_nodate} for speech. To fill this gap, we developed "NOIZEUS Birdsong" \cite{sainburg_2024_13947444}, a benchmark dataset modeled after NOIZEUS's methodology and structure.
We sampled recordings from 14 European starlings, with five 40-second songs from each bird, all recorded in an acoustically isolated chamber  \cite{arneodo2019acoustically}. 
To simulate realistic conditions, we added noise at four SNRs: 0, 5, 10, and 15 dB. Noise samples were taken from the "Soundscapes from around the world" dataset from Xeno Canto \cite{vellinga_2024}. We selected eight distinct soundscape categories which we named: "rain", "town", "wind", "waterfall", "insects", "swamp", "frogscape", and "forest". Each soundscape contains various sources of noise and were sampled from the European Starling's natural geographic range. The dataset exhibits diverse spectro-temporal noise characteristics, illustrated in \ref{sec:datasets} (Fig \ref{fig:birdsong-dataset}). 

\begin{figure}[!htbp]
\centering
\includegraphics[width=0.8 \textwidth]{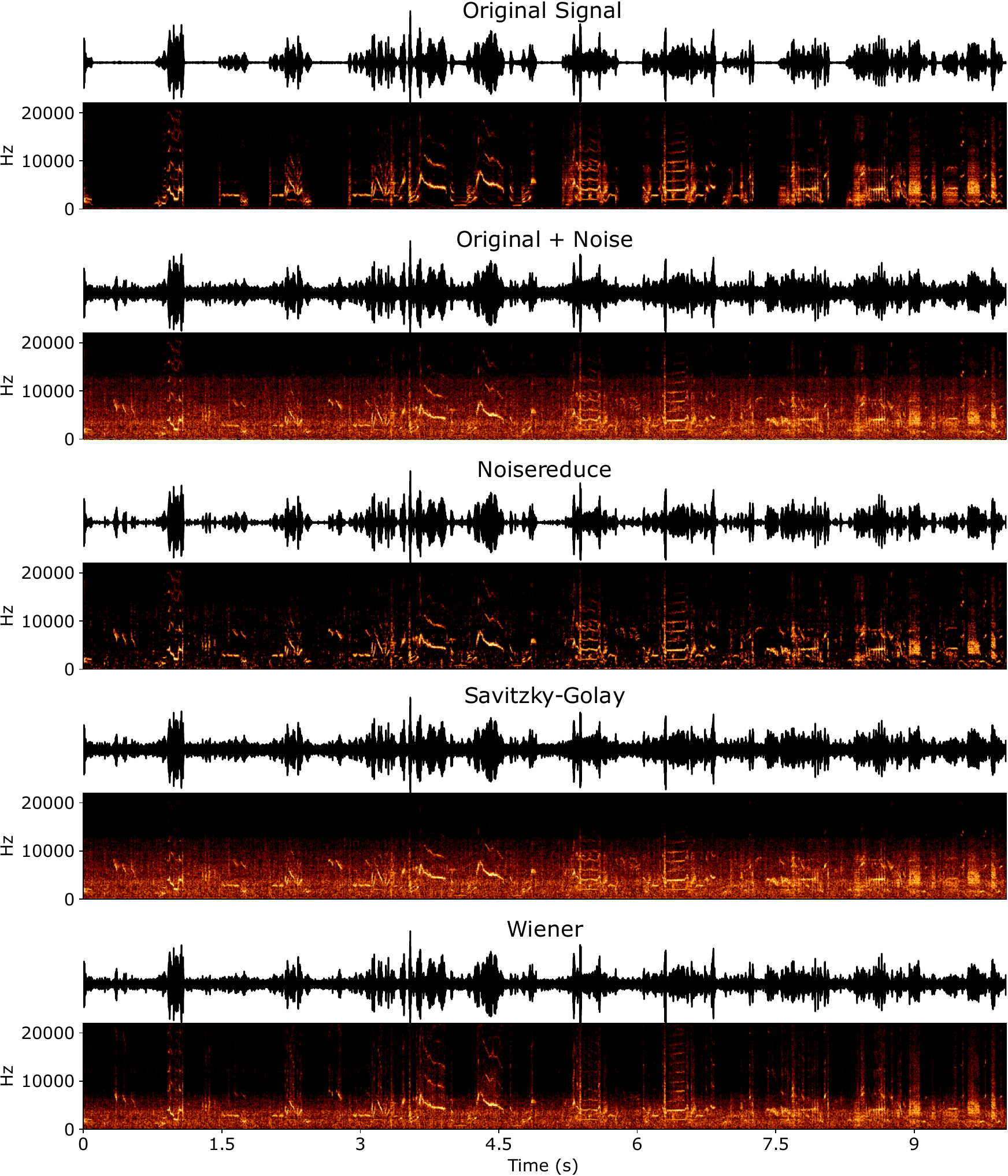}
\caption{
Noise reduction samples from different algorithms applied to the 'B335' sample from the NOIZEUS Birdsong dataset (SNR: 10 dB, waterfall noise).
}
\label{fig:birdsong-examples}
\end{figure}

We evaluated Noisereduce's performance using the NOIZEUS Birdsong dataset and compared it with other conventional noise reduction algorithms. We measure improvements in Segmental Signal-to-Noise Ratio (SegSNR), which evaluates the quality of noise reduction across temporal segments, and Source-to-Distortion Ratio (SDR), which quantifies both signal degradation and residual noise. We find that Noisereduce outperforms the other conventional algorithms on both metrics (Tables \ref{tab:segsnr}, \ref{tab:sdr}; Figure \ref{fig:birdsong-examples}).

%By calculating the statistical significance for each SNR level separately, we found that the differences between Noisereduce and other algorithms are statistically significant. For the SegSNR metric, Noisereduce significantly outperformed the Baseline at SNR levels of 0 dB, 5 dB, and 10 dB ($p<0.001$). At 15 dB SNR, Noisereduce's SegSNR score was $13.50 \pm 0.3913$ compared to the Baseline's $14.58 \pm 0.4014$; this difference was not statistically significant ($p=0.054$). For the SDR metric, Noisereduce demonstrated statistically significant improvements over the Baseline across all SNR levels ($p<0.001$).

\FloatBarrier

\begin{table}[ht]
    \small
    \centering
    \begin{tabular}{
        l 
        S[table-format=1.3(3)]
        S[table-format=1.3(3)]
        S[table-format=1.3(3)]
        S[table-format=1.3(3)]
    }
    \toprule
    {Algorithm} & {SNR 0} & {SNR 5} & {SNR 10} & {SNR 15} \\
    \midrule
    Baseline & -0.38 \pm 0.40 & 4.59 \pm 0.40 & 9.58 \pm 0.40 & \textbf{14.58 $\pm$ 0.40} \\
    Noisereduce (ours) & \textbf{6.96 $\pm$ 0.31} & \textbf{9.61 $\pm$ 0.33} & \textbf{11.78 $\pm$ 0.33} & 13.50 \pm 0.39 \\
    Savitzky-Golay & 0.27 \pm 0.48 & 4.45 \pm 0.46 & 8.41 \pm 0.43 & 11.91 \pm 2.47 \\
    Wiener & -0.09 \pm 0.42 & 4.51 \pm 0.41 & 8.61 \pm 0.39 & 11.82 \pm 0.37 \\
    \bottomrule
    \end{tabular}
    \caption{SegSNR [dB] performance metric on birdsong NOIZEUS dataset (mean $\pm$ SEM) for different algorithms across various SNR levels.}
    \label{tab:segsnr}
\end{table}

\begin{table}[ht]
    \small
    \centering
    \begin{tabular}{
        l 
        S[table-format=1.3(3)]
        S[table-format=1.3(3)]
        S[table-format=1.3(3)]
        S[table-format=1.3(3)]
    }
    \toprule
    {Algorithm} & {SNR 0} & {SNR 5} & {SNR 10} & {SNR 15} \\
    \midrule
    Baseline & -6.37 \pm 0.36 & -1.43 \pm 0.36 & 3.55 \pm 0.37 & 8.56 \pm 0.37 \\
    Noisereduce (ours) & \textbf{0.79 $\pm$ 0.40} & \textbf{4.94 $\pm$ 0.37} & \textbf{8.69 $\pm$ 0.31} & \textbf{11.61 $\pm$ 0.25} \\
    Savitzky-Golay & -5.41 \pm 0.49 & -0.93 \pm 0.49 & 3.61 \pm 0.47 & 7.95 \pm 0.44 \\
    Wiener & -5.71 \pm 0.43 & -0.71 \pm 0.44 & 4.18 \pm 0.45 & 8.53 \pm 0.38 \\
    \bottomrule
    \end{tabular}
    \caption{SDR [dB] performance metric on birdsong NOIZEUS dataset (mean $\pm$ SEM) for different algorithms across various SNR levels.}
    \label{tab:sdr}
\end{table}

\FloatBarrier

\subsection{Electrophysiology}
Extracellular electrophysiology is a key tool in recording single-neuron activity as animals interact with their environment. A challenge here is detecting extracellular spikes and assigning them to individual neurons, a process known as spikesorting. Current algorithms tackle this in steps: initially detecting spikes by thresholding amplitude or convolving the signal with spike templates, then iteratively clustering these putative spikes to estimate neuron identities, which provide templates for further detection. We tested whether Noisereduce could enhance initial spike detection by improving the SNR between spikes and background noise.

\begin{figure}[!htbp]
\centering
\includegraphics[width=0.64 \textwidth]{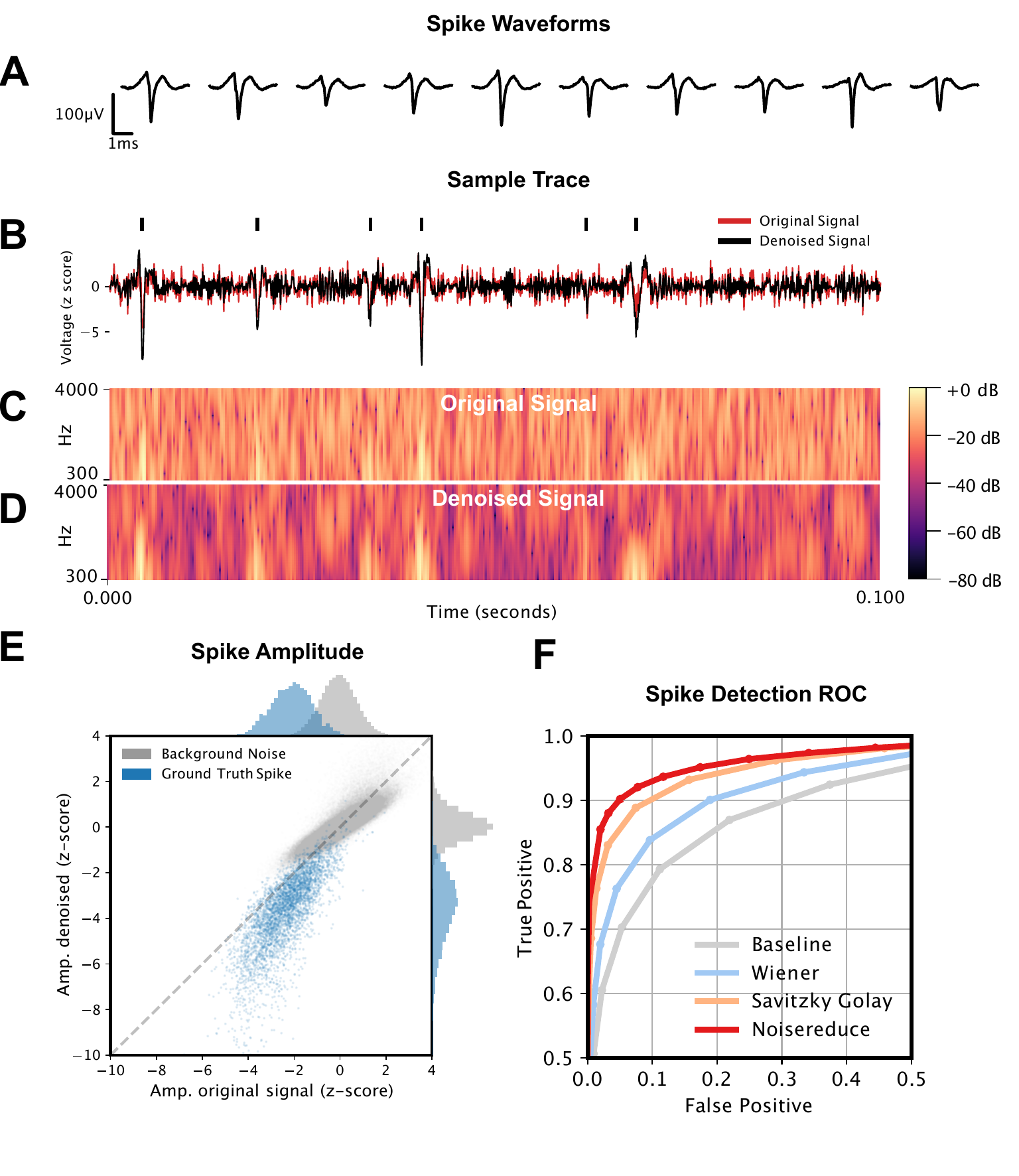}
\caption{
Noisereduce results on a simulated extracellular recording. (A) Sample neuron waveform templates. 
(B) A sample of 100ms of z-scored sampled neural data, with the original data in red and the denoised signal in black. 
(C-D) A spectrogram of the same data in B. 
(E) Amplitude of action potentials (blue) versus background noise (grey) in the original signal versus the denoised signal. 
(F) Reciever Operator Characteristic (ROC) curve of spike detection using the SpikeInterface \texttt{detect\_peaks} algorithm to detect spikes. 
}
\label{fig:neural-results}
\end{figure}

We created a dataset of biophysically realistic neural recordings using the MEArec library \cite{buccino2021mearec}, simulating extracellular electrophysiology. Ground truth spikes were simulated from 10 neurons (8 excitatory, 2 inhibitory, Fig \ref{fig:neural-results}A), with noise from 300 background neurons. Simulated data were used as real recordings lack ground truth.

We applied a modified Noisereduce approach to this data (Fig \ref{fig:neural-results}B), omitting the spectral mask smoothing step, which is computationally intensive and unnecessary for preliminary spike detection where spike shape is not used. We compared the output of Noisereduce to the untreated signal (bandpass filtered at 200-6000Hz; Fig \ref{fig:neural-results}C-D). We found that the spike amplitude (z-scored; Fig \ref{fig:neural-results}E) increased relative to background noise. To assess detection improvement, we used the SpikeInterface detection algorithm \cite{buccino2020spikeinterface} and computed an ROC curve by varying the detection threshold. Noisereduce was compared against three conditions: baseline bandpass filtering, Wiener filtering, and Savitzky-Golay filtering (Fig \ref{fig:neural-results}F). An Area Under the Curve (AUC) analysis found highest performance with Noisereduce (Noisereduce=0.97; Savitzky-Golay=0.96; Wiener = 0.94; Baseline=0.91), suggesting its suitability for initial spike detection.
Given that spectral masking can alter spike shapes, we advise using Noisereduce solely for initial spike detection, not clustering.

\FloatBarrier

\subsection{Seismology}
Seismic event detection methods focus on identifying the onset of these events, a critical step for accurately locating and characterizing seismic activity \cite{10.1785/BSSA0680051521, 10.1785/BSSA0840020366}. A widely used approach is the Short-Time Average over Long-Time Average (STA/LTA) algorithm \cite{10.1111/j.1365-246X.1990.tb00544.x, Trnkoczy2009UnderstandingAP}, which calculates the ratio of short-term to long-term signal averages to detect events. However, background noise from the environment and equipment makes detection less reliable, resulting in missed detections and false alarms. 
%To address these challenges, we modified the Noisereduce algorithm by removing the spectral mask smoothing step and setting a minimal hop length. This adjustment was needed to preserve the sharp signal transitions critical for onset detection.

\begin{figure}[!htbp]
\centering
\includegraphics[width=1.0 \textwidth]{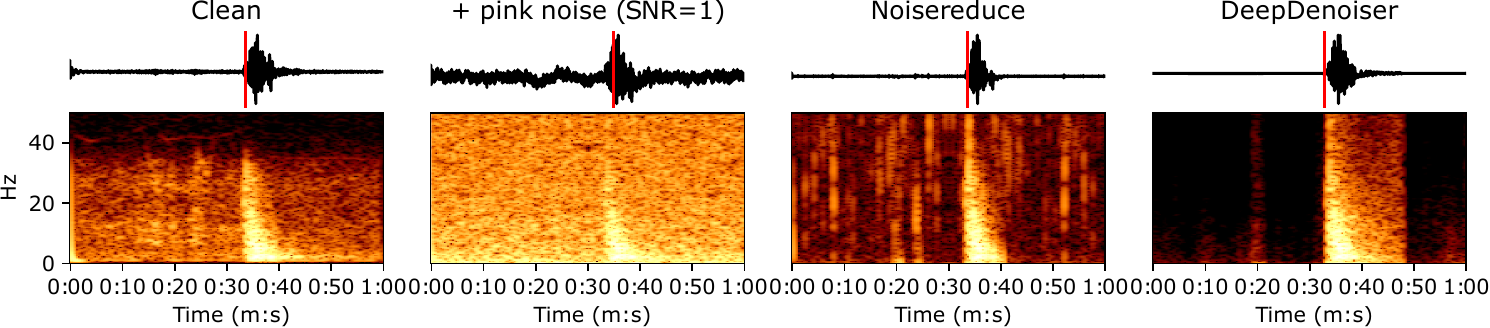}
\includegraphics[width=1.0 \textwidth]{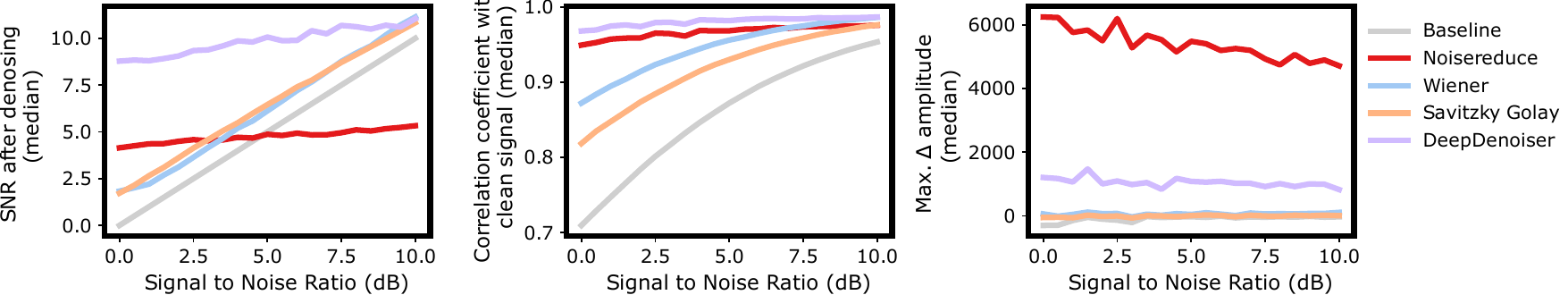}
\caption{
(Top) Seismic recording sample "ev0\_6.a01.gse2" from ObsPy dataset. The trigger, determined using the STA/LTA algorithm, marks the signal onset (red line). Noise added (pink, SNR = 1dB) and Noisereduce and DeepDenoiser are compared. 
(Bottom) Performance metrics for the seismology dataset. DeepDenoiser comparisons were generated using the DeepDenoiser API at "\url{https://ai4eps-deepdenoiser.hf.space}".
}
\label{fig:seismic-results}
\end{figure}

Following Zhu et al. (2019) \cite{8802278}, we tested Noisereduce on seismic waveforms from the ObsPy library \cite{10.1785/gssrl.81.3.530} (see Fig \ref{fig:seismic-results}). To simulate realistic conditions, we added white and pink noise at SNRs ranging from 0 to 15 dB and evaluated detection accuracy by comparing STA/LTA-detected onset times between denoised and clean recordings. As with the spike-detection analysis, we applied a modified Noisereduce, omitting the smoothing step. Noisereduce outperformed three baseline methods—no filtering, Wiener, and Savitzky-Golay—across all SNR levels, with the most significant improvements in low-SNR conditions (0 and 5 dB) for both noise types (see Tables \ref{tab:seismic-white} and \ref{tab:seismic-pink}).

\sisetup{
    detect-all,
    separate-uncertainty = true,
    table-format=1.3
}

\begin{table}[ht]
    \small
    \centering
    \begin{tabular}{
        l
        S[table-format=1.3(3)]
        S[table-format=1.3(3)]
        S[table-format=1.3(3)]
        S[table-format=1.3(3)]
    }
    \toprule
    {Algorithm} & {SNR 0} & {SNR 5} & {SNR 10} & {SNR 15} \\
    \midrule
    Baseline & 0.569 \pm 0.106 & 0.385 \pm 0.082 & 0.186 \pm 0.042 & 0.090 \pm 0.019 \\
    Noisereduce (ours) & \textbf{0.192 $\pm$ 0.065} & \textbf{0.187 $\pm$ 0.058} & \textbf{0.124 $\pm$ 0.034} & \textbf{0.069 $\pm$ 0.019} \\
    Wiener & 0.297 \pm 0.063 & 0.237 \pm 0.052 & 0.197 \pm 0.054 & 0.080 \pm 0.024 \\
    Savitzky-Golay & 0.242 \pm 0.058 & 0.201 \pm 0.045 & 0.169 \pm 0.044 & 0.090 \pm 0.023 \\
    % DeepDenoiser & 6.083 \pm 1.685 & 8.700 \pm 1.416 & 8.398 \pm 0.860 & 9.400 \pm 0.917 \\
    \bottomrule
    \end{tabular}
    \caption{Onset detection error (mean $\pm$ SEM) for different algorithms across various SNR levels for white noise.}
    \label{tab:seismic-white}
\end{table}

\begin{table}[ht]
    \small
    \centering
    \begin{tabular}{
        l
        S[table-format=1.3(3)]
        S[table-format=1.3(3)]
        S[table-format=1.3(3)]
        S[table-format=1.3(3)]
    }
    \toprule
    {Algorithm} & {SNR 0} & {SNR 5} & {SNR 10} & {SNR 15} \\
    \midrule
    Baseline & 0.392 \pm 0.102 & 0.308 \pm 0.073 & 0.226 \pm 0.052 & 0.111 \pm 0.028 \\
    Noisereduce (ours) & \textbf{0.129 $\pm$ 0.036} & \textbf{0.106 $\pm$ 0.025} & \textbf{0.074 $\pm$ 0.020} & \textbf{0.067 $\pm$ 0.017} \\
    Wiener & 0.253 \pm 0.085 & 0.244 \pm 0.057 & 0.193 \pm 0.039 & 0.093 \pm 0.027 \\
    Savitzky-Golay & 0.134 \pm 0.023 & 0.171 \pm 0.034 & 0.160 \pm 0.028 & 0.101 \pm 0.027 \\
    % DeepDenoiser & 8.755 \pm 1.809 & 6.635 \pm 1.746 & 8.302 \pm 0.817 & 9.740 \pm 0.660 \\
    \bottomrule
    \end{tabular}
    \caption{Onset detection error (mean $\pm$ SEM) for different algorithms across various SNR levels for pink noise.}
    \label{tab:seismic-pink}
\end{table}

\FloatBarrier

To further assess Noisereduce's quality in detecting seismic signals, we compared it to DeepDenoiser \cite{zhu2019seismic}, a deep neural network-based approach for denoising seismic waveforms.  We compared the SNR of the signal post-denoising, the correlation coefficient between clean and denoised signals, and the change in maximum amplitude of the signal from the clean recording. While Noisereduce underperforms compared to the deep learning approach on all metrics, its performance is closer to the deep learning model than any of the other conventional algorithms, with the exception of the amplitude of the denoised signal, which is more greatly decreased in Noisereduce (Fig. \ref{fig:seismic-results}).  

\FloatBarrier

\subsection{Run-Time Analysis}
\label{sec:runtime}
\begin{figure}[!htbp]
\centering
\includegraphics[width=0.7 \textwidth]{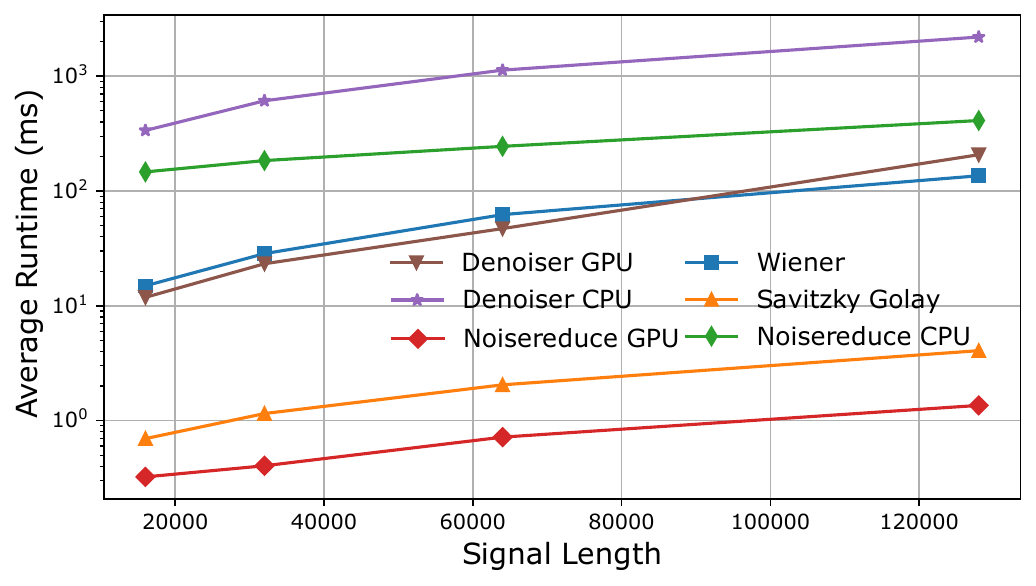}
\caption{
Runtime analysis comparing GPU-based Noisereduce, CPU-based Noisereduce, GPU-based Denoiser, CPU-based Denoiser, Wiener filter, and Savitzky-Golay filter on an RTX 3070 GPU with batch size of 32, and sample rate of 16 kHz.
}
\label{fig:runtime}
\end{figure}

Speed is a critical factor in selecting a noise reduction method, especially in applications requiring real-time or near-real-time analysis. Noise reduction algorithms are of limited use if they cannot process signals in a timely manner, as delays can become bottlenecks in analytical workflows.
Noisereduce supports GPU parallelization, which significantly improves processing speed.
To evaluate performance, we measured the average runtime across various signal lengths using an NVIDIA GeForce RTX 3070 GPU. The results (Fig \ref{fig:runtime}) demonstrate that GPU-accelerated Noisereduce outperforms other noise reduction algorithms, highlighting its potential for real-time applications. 
% For further details on the noise reduction algorithms evaluated, see  \ref{sec:algorithms}.

% In this section, we highlight the importance of speed in signal processing and provide a detailed comparison of the execution times of Noisereduce against other conventional algorithms.
% Noise reduction algorithms are of limited use if they cannot process signals in a timely manner, as delays can become bottlenecks in analytical workflows.
% , highlighting its potential for real-time applications.

% Noise reduction algorithms are of limited use if they cannot process signals in a timely manner, especially in applications requiring real-time or near real-time analysis.  In this section, we highlight the importance of speed in signal processing and provide a detailed comparison of the execution times of Noisereduce against other conventional algorithms. Fast processing not only enables immediate feedback and decision-making but also makes Noisereduce a suitable candidate for scenarios with limited computational resources. Thus, speed is a critical factor in selecting a noise reduction method, ensuring that the processing does not become a bottleneck in the flow of analytical tasks.

\FloatBarrier

\FloatBarrier

\section{Discussion}
In this work we provide a validation for Noisereduce as a domain-general noise-reduction algorithm. Our findings demonstrate that Noisereduce outperforms traditional noise reduction algorithms, making it suitable for a number of applications such as in bioacoustics and electrophysiology. Additionally, it can be used as a baseline comparison in domains where extensive domain-general machine-learning based approaches already exist.
An important advantage of Noisereduce is its support for GPU parallelization, which accelerates processing speed compared to CPU-only algorithms. The algorithm does not rely on training data, and its lightweight design makes it suitable for real-time use or resource-limited settings where deep learning may not be practical.
Noisereduce is publicly available as a Python package, actively maintained, and easy to use.

% In this work we provide a validation for Noisereduce as a domain-general noise-reduction algorithm. We observe that Noisereduce performs favorably over other traditional noise reduction algorithms, making it suitable for a number of applications such as in bioacoustics and electrophysiology, and can be used as a baseline comparison in domains where extensive domain-general machine-learning based approaches already exist. Noisereduce is publicly available as a Python package, actively maintained, and easy to use.

\bibliography{references}
\bibliographystyle{abbrvnat}

\section{Supplementary Information}

\subsection{Code Availability}
\label{sec:code}
The implementation of the Noisereduce algorithm is available at: \url{https://github.com/timsainb/noisereduce}. A future version of this software may be migrated to \url{https://github.com/noisereduce/noisereduce}. The experimental results, including all necessary configuration files and scripts for reproducing the experiments, are provided at: \url{https://github.com/noisereduce/paper_noisereduce}.

\subsection{Implementation Details}
\label{sec:implementation}

\subsubsection{Algorithm}

To reduce noise from time-series recordings, Noisereduce builds a mask over a time-frequency representation of the signal, which is used to mask noise from signal. The concept of spectral gating/subtraction/masking originates with Boll in 1979 \cite{boll1979suppression} and many variants of spectral gating have been developed since that time, ranging from simple statistics to more recent deep learning based approaches \cite{soni2018time}. Noisereduce generates a spectral mask by computing descriptive statistics over the time-frequency representation of noise clip and comparing them to the signal. The spectral mask is generated using the following steps:
\\

Compute a Short-Time Fourier Transform (STFT) over the time series noise clip (\(X_{\text{noise}}\)). If no noise clip is provided, treat the signal clip as the noise clip to compute statistics.

\begin{equation}
%X_n(f, t) = \sum_{m=0}^{N-1} X_{\text{noise}}[m + t \cdot h] \cdot w[m] \cdot e^{-j2\pi \frac{fm}{N}}
S_n = STFT(X_n, h, N)
\end{equation}

Where $h$ is the hop length hyperparameter, and N is the window length hyperparameter. \\

Convert the STFT (\(X_n\)) to decibels (\(X_{n\text{dB}}\))

\begin{equation}
X_{n\text{dB}}(f, t) = 20 \cdot \log_{10}(|X_n(f, t)| + \epsilon)
\end{equation}

Where $\epsilon$ is a small offset for stability. \\

We next compute descriptive statistics over the noise. For each frequency channel, compute the mean (\(\mu_n\)) and standard deviation (\(\sigma_n\)) over the noise spectrogram \(S_n\).

\begin{equation}
\mu_n(f) = \frac{1}{T} \sum_{t=1}^{T} X_{n\text{dB}}(f, t)
\end{equation}

\begin{equation}
\sigma_n(f) = \sqrt{\frac{1}{T} \sum_{t=1}^{T} \left(X_{n\text{dB}}(f, t) - \mu_n(f)\right)^2}
\end{equation}

where \( T \) is the total number of time frames.

These descriptive statistics are used to create a threshold for the mask. For each frequency channel, compute the noise threshold (\(\text{thresh}_n\)) from the mean (\(\mu_n\)), standard deviation (\(\sigma_n\)), and a hyperparameter (\(k\)) which sets the number of standard deviations above the mean to draw the threshold boundary.

\begin{equation}
\text{thresh}_n(f) = \mu_n(f) + k \cdot \sigma_n(f)
\end{equation}

We can then create the mask for the signal. Compute the STFT (\(S_X\)) of the signal clip (\(S\)).

\begin{equation}
%S_X(f, t) = \sum_{m=-\infty}^{\infty} S[m] \cdot w[m - t] \cdot e^{-j2\pi fm}
S_X = STFT(X, h, N)
\end{equation}

Convert the signal STFT (\(S_X\)) to decibels (\(S_{X\text{dB}}\)).

\begin{equation}
S_{X\text{dB}}(f, t) = 20 \cdot \log_{10}(|S_X(f, t)| + \epsilon)
\end{equation}

Compute a mask (\(M\)) over the signal spectrogram (\(S_{X\text{dB}}\)), based on the thresholds for each frequency channel.

\begin{equation}
M(f, t) = 
\begin{cases} 
1, & \text{if } S_{X\text{dB}}(f, t) > \text{thresh}_n(f) \\
0, & \text{otherwise}
\end{cases}
\end{equation}

Optionally, create a smoothing filter for the mask $SF$ in both frequency and time based on hyperparameters \(smooth_f\) and \(smooth_t\).

The smoothing filter \( SF(f, t) \) is defined as a separable matrix:

\begin{equation}
SF(f, t) = \frac{1}{Z} \cdot \left( L_f(f) \otimes L_t(t) \right)
\end{equation}

where \( \otimes \) denotes the outer product, and the components are defined as:

\begin{equation}
L_f(f) = 1 - \left| \frac{f - n_{\text{grad\_freq}}}{n_{\text{grad\_freq}}} \right|
\end{equation}

\begin{equation}
L_t(t) = 1 - \left| \frac{t - n_{\text{grad\_time}}}{n_{\text{grad\_time}}} \right|
\end{equation}

The expressions are defined for \( 0 \leq f \leq 2 \cdot n_{\text{grad\_freq}} \) and \( 0 \leq t \leq 2 \cdot n_{\text{grad\_time}} \), effectively creating symmetric triangular windows. The normalization constant \( Z \) is the sum of all elements in the smoothing matrix \( SF(f, t) \):

\begin{equation}
Z = \sum_{f} \sum_{t} \left( L_f(f) \otimes L_t(t) \right)
\end{equation}

\begin{equation}
SF(f, t) = \frac{1}{Z} \cdot W(f, t)
\end{equation}

where \( Z \) is the normalization constant such that \( \sum_{f,t} W(f, t) = 1 \).
\hfill \break

Apply the mask (\(M_{\text{smooth}}\)) to the STFT of the signal (\(S_X\)) by convolving \(M\) with \(S_X\) to produce the masked STFT (\(S_m\)).

\begin{equation}
%M_{\text{smooth}}(f, t) = \sum_{f'=-n_{\text{grad\_freq}}}^{n_{\text{grad\_freq}}} \sum_{t'=-n_{\text{grad\_time}}}^{n_{\text{grad\_time}}} M(f-f', t-t') \cdot SF(f', t')
% M_{\text{smooth}} = conv2d(S_X, SF)
S_m = conv2d(S_X, SF)
\end{equation}

Invert the masked STFT (\(S_m\)) back into the time-domain \(X_{\text{denoised}}\) using an inverse STFT.

\begin{equation}
%x_{\text{denoised}}[n] = \sum_{t} \sum_{f=0}^{N/2} S_m(f, t) \cdot e^{j2\pi \frac{fn}{N}} \cdot w[n - t \cdot h]
% x_{\text{denoised}} = iSTFT(S_m, h, N)
X_{\text{denoised}} = \text{iSTFT}(S_m, h, N)
\end{equation}

\paragraph{Non-stationary}
The non-stationary algorithm differs from the stationary version of Noisereduce in how the noise mask is computed. The central goal of the non-stationary algorithm is to compute a noise mask locally in time rather than globally across the entire recording or dataset, to account for fluctuations in the noise floor. To accomplish this, we simply compute the mean and standard deviation of the frequency components over a sliding window on $X$ without a noise clip, and then proceed with the rest of the algorithm normally. 

\paragraph{Soft Mask}
While the current implementation uses a binary mask (0 or 1), future work could explore a soft mask with values between 0 and 1 to achieve smoother signal-noise separation.

\subsubsection {Choosing Hyperparameters}
Noisereduce relies on a small number of hyperparameters which will need to be set to match your data. 
\\

The main two parameters to consider are \texttt{n\_std\_thresh\_stationary} and \texttt{prop\_decrease}. \texttt{n\_std\_thresh\_stationary} sets the threshold for what to consider signal in terms of standard deviations of power above (or below) the mean power for each frequency channel. \texttt{prop\_decrease} then determines the extent to which we remove the below-threshold noise. 
We additionally include \texttt{noise\_window\_size\_nonstationary\_ms} in the nonstationary version of the algorithm, which is the window over which threshold statistics are computed. 

\texttt{freq\_mask\_smooth\_hz} and \texttt{time\_mask\_smooth\_ms}, are used to smooth the mask using a Gaussian kernel, with the shape of the kernel defined by those parameters. 
\\ 

Finally, \texttt{n\_fft}, \texttt{win\_length}, and  \texttt{hop\_length} are all parameters used to compute the spectrogram and should be set at values that would visibly capture spectrotemporal structure in your signal, if you were to plot the spectrogram. 

\FloatBarrier
\begin{table}[h]
\centering
\small % Change this to \footnotesize or \scriptsize for even smaller text
\renewcommand{\arraystretch}{1.5} % Adjust vertical padding
\setlength{\tabcolsep}{0.5em} % Adjust horizontal padding
\begin{tabular}{|l|p{8cm}|}
\hline
\textbf{Parameter} & \textbf{Description} \\ \hline
n\_fft & 
Length of the windowed signal after padding with zeros, by default 1024. \\ \hline
win\_length & 
Each frame of audio is windowed by ``window`` of length ``win\_length`` and then padded with zeros to match ``n\_fft``, by default None. \\ \hline
hop\_length & 
Number of audio samples between adjacent STFT columns, by default None. \\ \hline
n\_std\_thresh & 
Number of standard deviations above mean to place the threshold between signal and noise, by default 1.5. \\ \hline
noise\_window\_size\_nonstationary\_ms & 
The window size (in milliseconds) to compute the noise floor over in the non-stationary algorithm, by default 1. \\ \hline
freq\_mask\_smooth\_hz& 
The frequency range to smooth the mask over in Hz, by default 500. \\ \hline
time\_mask\_smooth\_ms & 
The time range to smooth the mask over in milliseconds, by default 50. \\ \hline
prop\_decrease & 
The proportion to reduce the noise by (1.0 = 100\%), by default 1.0. \\ \hline
\end{tabular}
\caption{Parameters for noise reduction}
\label{tab:parameters}
\end{table}
\FloatBarrier

\subsection{Datasets}
\label{sec:datasets}

\paragraph{Speech}
The evaluation included thirty phonetically balanced speech utterances from the "\textit{Noisy Speech Corpus}" (NOIZEUS) database \cite{noauthor_noizeus:_nodate, yi_hu_subjective_2006, hu_evaluation_2008}, a database specifically designed for noisy speech research. These utterances were combined with eight distinct real-world noise types, including suburban train, babble, car, exhibition hall, restaurant, street, airport, and train station noises, at SNRs of $0$ dB, $5$ dB, $10$ dB, and $15$ dB, following Method B of the ITU-T P.56 standard \cite{noauthor_p.56_nodate}.

\begin{figure}[!htbp]
\centering
\includegraphics[width=0.9 \textwidth]{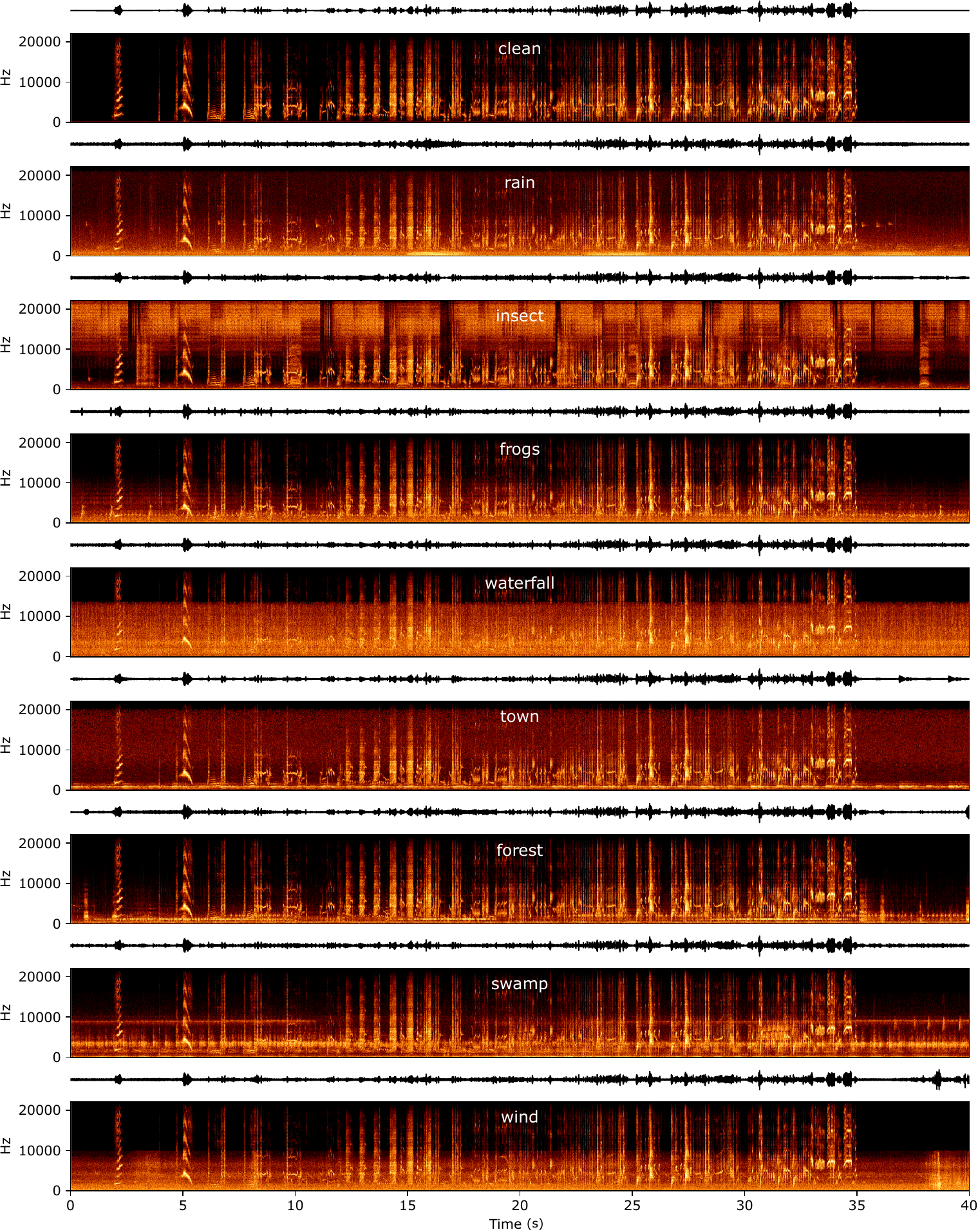}
\caption{
Spectrograms of a sample from the "Birdsong NOIZEUS" dataset at an SNR of 10 dB, showcasing the clean signal, the noisy signals with different types of environmental noise. 
}
\label{fig:birdsong-dataset}
\end{figure}

\paragraph{Birdsong}
We created the Birdsong NOIZEUS dataset \cite{sainburg_2024_13947444} in the likeness of the speech NOIZEUS dataset \cite{noauthor_noizeus:_nodate}. We selected 70 song samples from 14 European starlings (5 samples of 40 seconds each). These recordings were selected from a larger collection previously gathered by the authors for prior publications \cite{arneodo2019acoustically, sainburg2019parallels}.
The original dataset contains several hundred 30 to 60 second recordings per bird, obtained from wild-caught European starlings in Southern California. Recordings were performed in acoustically isolated chambers to ensure high-quality audio capture.
We sampled noise from the "Soundscapes from around the world" dataset from Xeno Canto \cite{vellinga_2024}. We hand selected 8 soundscapes from this dataset which we named "rain", "town", "wind", "waterfall", "insects", "swamp", "frogscape", and "forest". Each soundscape contains various sources of noise and were sampled from the European Starling's natural range. For each song, we selected a different segment of the soundscape (soundscapes were around 5-20 minutes each). An example of the dataset can be seen in Figue \ref{fig:birdsong-dataset}. We set the SNR based on loudness measured using the pyloudnorm Python library \cite{steinmetz2021pyloudnorm}. Additionally, for each song and noise clip we included a 1-second clip of noise sampled randomly (at the same SNR of the audioclip). This dataset is publicly available on Zenodo (DOI: 10.5281/zenodo.13947444).

\paragraph{Seismology}
The seismic data used in this study was obtained from the ObsPy Trigger/Picker Tutorial \cite{10.1785/gssrl.81.3.530}, including  waveform recordings from three seismic stations: EV, RJOB, and MANZ. The dataset, recorded in January 1970, contains natural seismic events specifically selected for evaluating triggering and picking algorithms. For our analysis, we used all available signals from the trigger dataset, excluding those with low SNR. To simulate realistic seismic and instrumental noise, we added both white noise and pink noise at varying SNR ratios (0, 5, 10, and 15 dB).

\paragraph{Electrophysiology}
Electrophysiology datasets were generated using the MEArec \cite{buccino2021mearec} Python library so that we would have access to ground truth spiking events alongside electrophysiology. Some non-simulated ephys datasets record ground truth events, e.g. by pairing extracellular recordings with intracellular recordings \cite{magland2020spikeforest}, but, since not all cells are recorded intracellularly, they are of limited value in differentiating between false positive and true positive detections of other cells. facilitates the generation of customizable extracellular spiking activity datasets by leveraging biophysically detailed simulations. It achieves this by first creating templates of extracellular action potentials using realistic cell models, positioned around electrode probes within a simulation environment. These cell models, drawn from established neuroscience databases, undergo intracellular simulation to compute transmembrane currents using tools like NEURON, while the extracellular potentials are calculated using methods such as the line-source approximation via the LFPy package. This process allows MEArec to accurately simulate various neural dynamics and probe configurations, offering a flexible framework for evaluating and developing spike sorting methods under controlled experimental conditions.

We generated a dataset comprising a monotrode (single channel) recording 10 minutes in length. The recording had 10 neurons (8 excitatory and 2 inhibitory). Spikes ranged in amplitude from 75-150uV. Background noise was generated using 300 simulated neurons that were further away from the probe (each with a maximum amplitude of 75uV). Simulated data were bandpass filtered between 300 and 6000 Hz.

\subsection{Additional Algorithms}
\label{sec:algorithms}
\paragraph{Wiener Filter}
The Wiener filter, as implemented in \cite{lim1990}, is an adaptive noise reduction algorithm that analyze local statistics within a sliding window. The filter adapts to local signal characteristics by weighting the difference between the noisy observation and the local mean based on the local variance. 
The filter applies minimal smoothing in high-variance regions to preserve significant signal features, while employing more aggressive smoothing in low-variance areas presumed to be noise-dominated.

The output signal $\widehat x\left[ n \right]$ at index $n$ is computed using:
\begin{equation}
    \widehat x\left[ n \right] = {\mu _n} + \frac{{{\sigma ^2}_n - {\nu ^2}}}{{{\sigma ^2}_n}}\left( {y\left[ n \right] - {\mu _n}} \right)
\end{equation}

where $y[n]$ is the observed noisy signal, $\mu_n$ is the local mean within a window centered around $n$, $\sigma_n^2$ is the variance of the signal in that window, and $\nu^2$ is the estimated noise variance calculated as the average of all local variances across the signal.

\paragraph{Iterative Wiener}
The Iterative Wiener \cite{1163086} performs noise reduction in the frequency domain by iteratively computing a Wiener filter for each frame. The Wiener filter for each frequency $\omega$ is defined as:

\begin{equation} H(\omega) = \frac{P_x(\omega)}{P_x(\omega) + P_n(\omega)} \end{equation}

where $P_x(\omega)$ is the speech power spectral density, and $P_n(\omega)$ is the noise variance. 

To determine if a frame contains speech, a simple energy threshold is used. When speech is detected, the algorithm refines the clean signal estimate by iteratively calculating Linear Predictive Coding (LPC) coefficients of the input frame, which model the vocal tract as an all-pole filter. These LPC coefficients help estimate the speech power spectrum and update the Wiener filter to reduce noise. After denoising, new LPC coefficients are calculated from the denoised signal, further improving the filter.

When no speech is detected, the noise variance is updated. The algorithm uses an IIR filter to smooth the noise estimate. The noise variance ${P_n(\omega)}$ is updated as follows:

\begin{equation} P_n(\omega) = \alpha P_{n-1}(\omega) + (1 - \alpha) E_{\text{input frame}} \end{equation}

where $\alpha$ is the smoothing factor, and $E_{\text{input frame}}$ is the energy of the input frame. 

\paragraph{Savitzky-Golay Filter}
The Savitzky-Golay filter \cite{savitzky1964} is a smoothing technique that employs a polynomial fitting approach. By fitting low-degree polynomials to successive subsets of adjacent data points, it effectively reduces noise while preserving important signal features. The output signal $\widehat x\left[ n \right]$ at index $n$ is computed using:
\begin{equation}
    \widehat x\left[ n \right] = \sum\limits_{j =  - m}^m {{c_j}{y_{i + j}}}
\end{equation}

where $2m + 1$ represents the window size, ${y_{i + j}}$ are the input data points within the window, and $c_j$ are the convolution coefficients derived by the polynomial fitting.

\paragraph{Spectral Subtraction}
The Spectral Subtraction algorithm \cite{boll1979} performs noise reduction by subtracting an estimate of the noise spectrum from the spectrum of the noisy signal. It operates under the assumption that the noise is additive and uncorrelated with the signal. The output signal spectrum $\widehat X\left( w \right)$ is computed as:
\begin{equation}
    \left| {\widehat X\left( \omega  \right)} \right| = \max \left( {0,\,\,\left| {Y\left( \omega  \right)} \right| - \left| {\widehat N\left( \omega  \right)} \right|} \right)
\end{equation}

where $Y(\omega)$ is the Fourier transform of the noisy signal, and $\widehat{N}(\omega)$ is the estimated noise spectrum. The $\max$ function ensures that the resulting magnitude is non-negative.
The noise spectrum $\widehat{N}(\omega)$ is obtained during periods of silence in the signal, under the assumption that only noise is present.
% ASK TIM HOW HE ESTIMATED THE NOISY SPECTRUM

To reconstruct the clean signal, the inverse Fourier transform is applied to $\widehat{X}(\omega)$ while using the phase information from the noisy signal $Y(\omega)$.

\paragraph{Subspace}
The Subspace algorithm \cite{397090, 5743782} performs noise reduction by projecting the noisy signal $y[n]$ onto a lower-dimensional subspace that primarily contains the clean signal, while the noise is assumed to be in the complementary subspace. 

An eigendecomposition is performed on the matrix,
\begin{equation}
    \Sigma  = {R_n}^{ - 1}{R_y} - I
\end{equation}
where $R_n$ is the noise covariance matrix, $R_y$ is the covariance matrix of the input noisy signal, and $I$ is the identity matrix. $R_n$ is obtained during periods of silence in the signal, under the assumption that only noise is present. 

The cleaned signal $\widehat x[n]$ is obtained by projecting the noisy signal $y[n]$ onto the signal subspace,
\begin{equation} 
    \widehat x[n] = \mathbf{H}_{opt} \cdot y[n] 
\end{equation}

where $\mathbf{H}_{opt}$ is the projection matrix, derived from the positive eigenvectors of $\Sigma$.

\paragraph{DeepDenoiser} 
The DeepDenoiser \cite{8802278} is a deep neural network designed to denoise seismic signals by learning to separate the signal from noise.
It is trained on datasets containing both noisy and clean waveform data. During the denoising process, the input seismic signal is first converted into the time-frequency domain. The network then uses a series of fully convolutional layers with skip connections to generate two masks: one for the signal and one for the noise. Finally, the denoised signal and the estimated noise are obtained by applying the inverse Short Time Fourier Transform.

\paragraph{Denoiser} 
Denoiser \cite{defossez2020realtimespeechenhancement} is a deep learning model designed to denoise speech signals. It processes raw audio waveforms using an encoder-decoder architecture with skip connections. The model is optimized across both time and frequency domains through multiple loss functions and is trained end-to-end on paired datasets of noisy and clean speech.

\subsection{Metrics}
\label{sec:metrics}
\paragraph{STOI}
Short-Time Objective Intelligibility (STOI) \cite{taal_short-time_2010} is a widely used objective metric for assessing speech intelligibility, particularly in noisy environments. It functions by comparing short-time segments of clean reference and degraded speech signals, quantifying the level of degradation in terms of intelligibility. STOI calculates the correlation between the two signals over overlapping time windows, producing a score between 0 and 1. Higher scores indicate better intelligibility.

\paragraph{PESQ}
Perceptual Evaluation of Speech Quality (PESQ) \cite{941023} is another objective metric designed to evaluate speech quality as perceived by human listeners. It incorporates a psychoacoustic model to simulate the human auditory system, comparing degraded or processed speech to a clean reference. PESQ captures both time-domain distortions and perceptual differences, generating a score between -0.5 and 4.5. Higher scores signify better perceived speech quality. Unlike STOI, which focuses on intelligibility, PESQ is more concerned with overall speech quality.

\paragraph{SDR}
Source Distortion Ratio (SDR) evaluates the quality of source separation by measuring the logarithmic ratio of the power of the target source signal to the power of distortions, such as interference, noise, and artifacts. Higher SDR values indicate better separation performance, with fewer distortions. SDR is defined as follows:
\begin{equation}
    SDR = 10{\log _{10}}\left( {\frac{{{{\left| x \right|}^2}}}{{{{\left| {x - \widehat x} \right|}^2}}}} \right)
\end{equation}
where $x$ is the true clean signal, and $\widehat x$ is the estimated signal.

\paragraph{SegSNR}
Segmental Signal-to-Noise Ratio (SegSNR or SSNR), a modified version of Signal-to-Noise Ratio (SNR), provides a more localized assessment of signal quality. While traditional SNR evaluates signal-to-noise ratios across the entire signal, SegSNR calculates SNR within smaller segments and then averages these values. SegSNR is defined as follows:
\begin{equation}
    SegSNR = \frac{1}{N}\sum\limits_{n = 1}^N {SNR\left( n \right)} 
\end{equation}
where $N$ is the total number of segments, and $SNR(n)$ is the SNR for the n-th segment.

\paragraph{AUC}
The Receiver Operating Characteristic (ROC) curve is used to evaluate the performance of binary classifiers. The ROC curve plots true positive rate against false positive rate for various classification thresholds. The Area Under the Curve (AUC) quantifies the overall performance, ranging from $0.5$ (random guessing) to $1.0$ (perfect classification).

\end{document}